\documentclass[12pt,preprint]{aastex}

\shorttitle{MOST Direct Imaging}
\shortauthors{Rowe et al.}

\begin{document}

\title{An Upper Limit on the Albedo of HD 209458b:\\
Direct Imaging Photometry with the MOST Satellite\footnote{MOST
is a Canadian Space Agency mission, operated jointly by Dynacon,
Inc., and the Universities of Toronto and British Columbia, with
assistance from the University of Vienna.}}

\author{Jason F. Rowe, Jaymie M. Matthews}

\affil{University of British Columbia\\
6224 Agricultural Road\\ 
Vancouver BC, V6T 1Z1\\
rowe,matthews@astro.ubc.ca}

\author{Sara Seager}

\affil{Carnegie Department of Terrestrial Magnetism\\
Washington, D.C.}

\author{Rainer Kuschnig}
\affil{Department of Physics and Astronomy, University of British Columbia \\ 
6224 Agricultural Road, Vancouver, BC V6T 1Z1, Canada}

\author{David B. Guenther}
\affil{Department of Astronomy and Physics, St. Mary's University\\
Halifax, NS B3H 3C3, Canada}

\author{Anthony F.J. Moffat}
\affil{D\'epartement de physique, Universit\'e de Montr\'eal \\ 
C.P.\ 6128, Succ.\ Centre-Ville, Montr\'eal, QC H3C 3J7, Canada}

\author{Slavek M. Rucinski}
\affil{David Dunlap Observatory, University of Toronto \\
P.O.~Box 360, Richmond Hill, ON L4C 4Y6, Canada}

\author{Dimitar Sasselov}
\affil{Harvard-Smithsonian Center for Astrophysics \\ 
60 Garden Street, Cambridge, MA 02138, USA}

\author{Gordon A.H. Walker} 
\affil{Department of Physics and Astronomy, University of British Columbia \\ 
6224 Agricultural Road, Vancouver, BC V6T 1Z1, Canada}

\author{Werner W. Weiss}
\affil{Institut f\"ur Astronomie, Universit\"at Wien \\ 
T\"urkenschanzstrasse 17, A--1180 Wien, Austria}

\begin{abstract}
We present space-based photometry of the transiting exoplanetary
system HD 209458 obtained with the MOST (\textit{Microvariablity and
Oscillations of STars}) satellite, spanning 14 days and covering 4
transits and 4 secondary eclipses. The HD 209458 photometry was
obtained in MOST's lower-precision Direct Imaging mode, which is used
for targets in the brightness range $6.5 \geq V \geq 13$.  We describe
the photometric reduction techniques for this mode of observing, in
particular the corrections for stray Earthshine.  We do not detect the
secondary eclipse in the MOST data, to a limit in depth of 0.053 mmag
($1\sigma$). We set a 1$\sigma$ upper limit on the planet-star flux
ratio of $4.88 \times 10^{-5}$ corresponding to a geometric albedo
upper limit in the MOST bandpass (400 to 700 nm) of 0.25.  The
corresponding numbers at the $3\sigma$ level are $1.34 \times 10^{-4}$
and 0.68 respectively. HD 209458b is half as bright as Jupiter in the
MOST bandpass. This low geometric albedo value is an important
constraint for theoretical models of the HD209458b atmosphere, in
particular ruling out the presence of reflective clouds.  A second
MOST campaign on HD 209458 is expected to be sensitive to an exoplanet
albedo as low as 0.13 (1$\sigma$), if the star does not become more
intrinsically variable in the meantime.
\end{abstract}


\section{Introduction}\label{intro}

Since the discovery of the giant planet orbiting 51 Pegasi
\citep{may95} a decade ago, and the subsequent detection of about 160   
exoplanetary systems around solar-type stars \citep{sch05}, 
the task of acquiring information about the atmospheres of
giant close-in planets has been particularly challenging.
Observations of the transits of the  
exoplanet HD 209458b have revealed the presence of sodium in its 
atmosphere \citep{cha02} and a hydrogen cloud around it \citep{vid03}.
The Spitzer infrared space observatory has detected the eclipse 
of HD 209458b in its thermal emission at 24-$\mu$ \citep{dem05a}, 
yielding a brightness temperature at that wavelength of $1130 \pm 150$ 
K.

We report here an attempt to detect the eclipse of HD 209458b (orbital
period $\sim$ 3.5 d) in 
reflected light at optical wavelengths, with photometry from the MOST
(\textit{Microvariablity and Oscillations of STars}) satellite
\citep{wal03,mat04}.  The reflected light 
signal from an exoplanet is sensitive to the composition of its 
atmosphere, as well as the nature and filling factor of cloud particles
suspended in that atmosphere \citep{sea00,gre03,bur05}.
The proportion of scattered to absorbed radiation is critical
to the planet's energy balance and hence an albedo measurement is key
to understanding its atmosphere.

The MOST satellite houses a 15-cm optical telescope feeding a CCD 
photometer through a single broadband filter (350 - 750 nm), which is
capable of sampling target stars up to 10 times per minute.  From the
vantage point of its 820-km-high circular Sun-synchronous polar orbit
with a period of 101.413 minutes,
MOST can monitor stars nearly continuously for up to 8 weeks in a 
Continuous Viewing Zone (CVZ) with a declination range of $+36^{\circ} 
\geq \delta \geq -18^{\circ}$.  The satellite was designed to achieve 
photometric precision of a few parts per million (ppm $\sim \mu$mag) at 
frequencies ($>$ 1 mHz) in the Fourier domain.  The necessary 
photometric stability is achieved by projecting an extended image of the 
instrument pupil illuminated by the Primary Science Target via one of 
an array of Fabry microlenses above the MOST Science CCD (a 1K $\times$ 
1K E2V 47-20 detector).  For exposure times up to 1 minute long, and 
observing runs of at least 1 month, this Fabry Imaging mode can reach
the desired precision of about 1 ppm at frequencies greater than 1 mHz
for stars brighter than V $\sim$ 6.5.

Fainter stars can be monitored (simultaneously, or independently of the
Fabry Imaging mode) in an open area of the Science CCD not covered by 
the $6 \times 6$ Fabry microlens array and its chromium field stop mask.
In the Direct Imaging field, defocused star images are projected, with
a Full-Width Half-Maximum (FWHM) of about 2.5 pixels.  (The focal plane
scale of MOST is about 3 arcsec/pixel.)  The photometric precision 
possible in the Direct Imaging Mode is not as good as for the Fabry 
photometry as Direct Imaging targets are fainter and more sensitive to
CCD calibrations.  However, the unprecedented duty cycle of the MOST
observations 
and the thermal/mechanical stability of the instrument yield impressive 
results.  The point-to-point precision of the photometry reported here
is about 3 millimag for a 1.5s exposure during low stray light
conditions, with sampling rates as high as 10 exposures per minute and 
a duty cycle of 85\% in 14 days of observation, where a duty cycle of
100\% represents no loss of data acquisition.

Photometry of this quality and coverage represents a unique opportunity 
to explore the HD 209458 transiting system.  The star is not too bright
for the MOST Direct Imaging mode ($V = 7.65$) and it is well placed 
in the MOST CVZ, observable for up to about a month and a half.  MOST 
data have many applications to this system:  (1) accurate transit timing 
to refine the near-zero orbital eccentricity and check for effects of 
orbital precession; (2) studies of the shapes of transit ingress and 
egress to search for large moons around HD 209458b; (3) searching for 
transits of other smaller planets in the systems with different orbital 
periods, as predicted by some models to explain the dynamical stability 
of HD 209458b (e.g., \citet{ida04}); (4) revealing subtle intrinsic 
variability in the star HD 209458a and possible interactions with the 
close-in planet; and (5) detection of the eclipse of HD 209458b in 
optical light to directly measure the geometric albedo of the planet.

In this paper, we report on MOST observations and analysis for 
eclipse detection in this system.  In the next two sections, we describe 
the data and the MOST Direct Imaging reduction scheme, as a reference 
for MOST Direct Imaging results published here and elsewhere.  We then 
present the reduced photometry of HD 209458, the eclipse analysis, and 
the upper limit on the depth of the eclipse.  We translate this into an 
upper limit on the optical albedo of the exoplanet, and discuss its 
implications in atmosphere and cloud models.  Finally, we predict the 
impact of future planned MOST observations of HD 209458 and the potential 
for other known transiting exoplanet systems.

\section{MOST Direct Imaging Photometry}

MOST is a microsatellite (mass = 54 kg; peak solar power = 39 W) with 
limited onboard processing capability, memory, and downlink.  Hence it
is not possible to transfer the entire set of $1024 \times 1024$ pixels 
of the Science CCD to Earth at a rapid sampling rate and with an ADC 
(Analogue-to-Digital Conversion) of 14 bits (necessary to preserve 
variability information at the ppm level).  Small segments of the CCD 
("subrasters") are stored, which contain key portions of the target 
field.  This usually includes the Primary Science Target Fabry Image, 
7 adjacent Sky Background Fabry Images, and subrasters for dark and bias 
readings.

\subsection{Data Format}

There is also the option to sample several nearby Secondary Science 
Targets in the Direct Imaging field (less than about $0.5^{\circ}$ 
away), by placing subrasters of typical dimensions $20 \times 20$ pixels 
over those stars. These targets are automatically monitored with the 
same sampling as the Primary Fabry Imaging target.  The Fabry Image 
illuminates about 1500 pixels, while each Direct Image illuminates a
PSF out to a radius of several pixels.  When simultaneously observing
Fabry and Direct Imaging targets to avoid saturation, the
Direct Imaging Targets must be at least  
5.5 magnitudes fainter than the Fabry Target.  For example, for MOST's 
first Primary Science Target, Procyon, the $V = 8$ star HD 61199 was 
chosen for the Direct Imaging field and was discovered to be a 
multiperiodic $\delta$ Scuti pulsator (see Fig. 2c in \citet{mat04})
with a peak amplitude of about 1 millimag .

It is also possible to select a star as bright as $V = 6.5$ as the
principal science target in the Direct Imaging field, without a brighter 
Fabry target.  Then the exposure time and sampling rate can be optimized 
for the Direct Imaging target.  This was the case for HD 209458 (V=7.7).

\begin{figure}
\begin{center}
\plotone{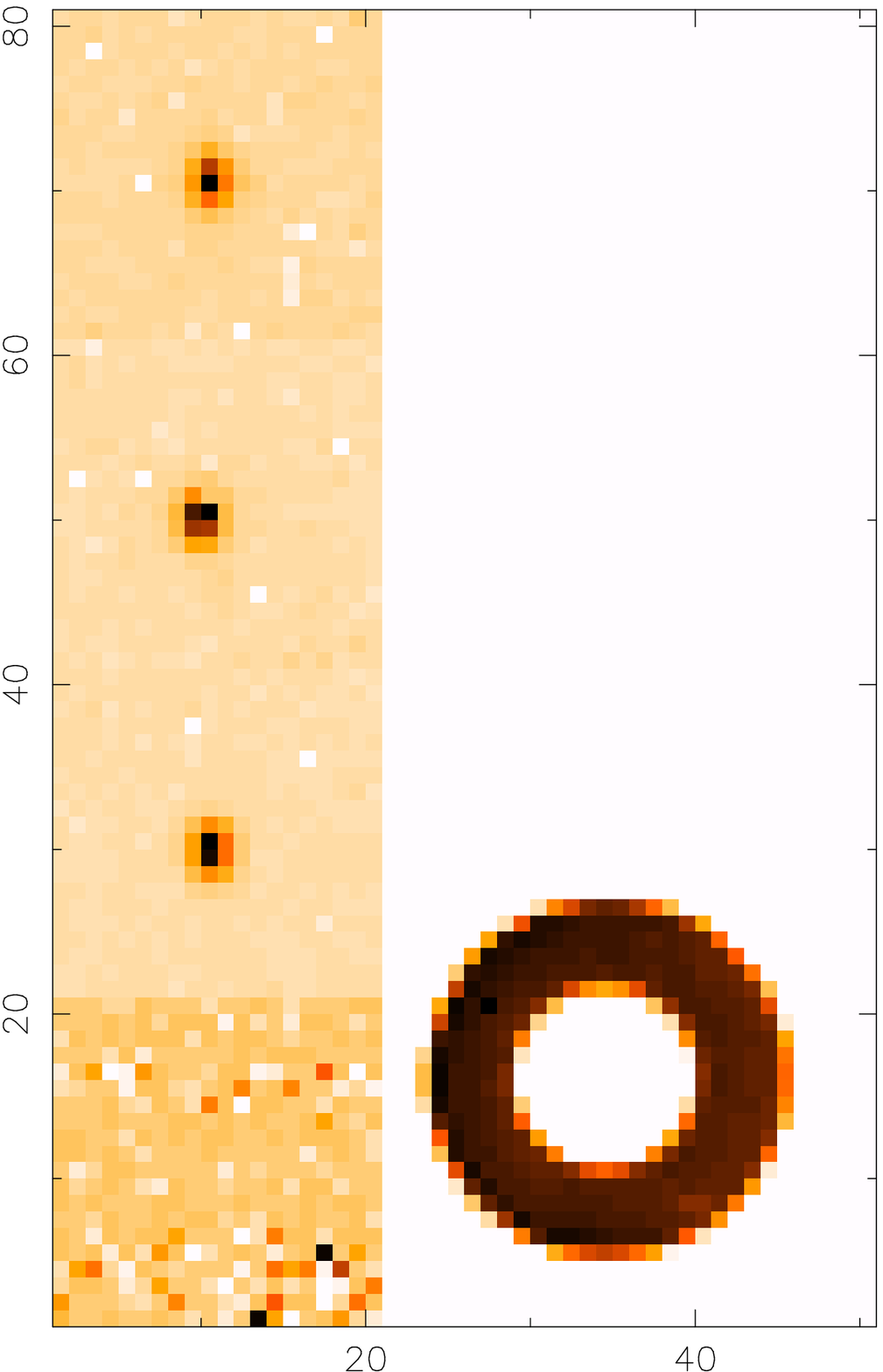}
\end{center}
\caption{The FITS file layout is shown.  There are 5 subrasters
  visible in this example, 3 containing stellar direct images, the
  Fabry projection and region shielded from light.  The axis units are
  in pixels.  The Fabry image has been binned 2x2.}
\label{fig:ccdlayout}
\end{figure}

When the binary data stream is transfered from the satellite it is
converted to a FITS format image. The layout is shown in Figure
\ref{fig:ccdlayout}.  The locations of each subraster in the MOST 
Science CCD (in x-y pixel coordinates) are specified in the FITS file 
header.  Typically, the FITS file contains resolved subrasters for the 
Fabry image, 2-4 Direct images and 1 dark. The header also contains 
the pixel sums for various bias and dark regions, as well as satellite
and instrument telemetry (e.g., spacecraft attitude control data; 
CCD focal plane temperatures) to allow additional photometric 
calibrations on the ground.  The data format is unique to MOST, which 
necessitated the development of custom software to handle and process
MOST data.  Examples of MOST raw data and a document describing the 
FITS file and header format in detail are available in the MOST Public
Data Archive at the MOST Mission website: {\sf www.astro.ubc.ca/MOST}.

\subsection{Photometric Reduction}

In general, the approach to reducing MOST Direct Imaging photometry is 
similar to groundbased CCD photometry, applying traditional aperture and 
PSF (Point-Spread Function) approaches to the subrasters.  However, there
are several aspects and challenges specific to MOST data.  For example, 
the MOST instrument has no on-board calibration lamp for correction of 
pixel-to-pixel sensitivity variations ("flatfielding").  In its orbit, 
MOST passes through a region of enhanced cosmic ray flux known as the 
South Atlantic Anomaly (SAA).  There are also phases of increased stray 
light from scattered Earthshine which repeat regularly during each 
satellite orbit ($P = 101$ min).

\subsubsection{Dark Correction\label{dark}}

To lower cost and increase reliability, the MOST instrument has no moving
parts, so there is no mechanical shutter which can cut off light to the 
entire CCD for dark measurements; exposures are ended by rapid charge 
transfer into a frame buffer on the CCD.  Dark measurements are obtained 
from portions of the CCD shielded from light by a chromium mask above the 
focal plane.  

One-dimensional dark current correction is done by using averages from 
these dark subrasters; the averages are computed on the satellite and the 
full raster information is not available on the ground (to meet downlink 
limitations while maximizing the amount of stellar data available). The 
values for each dark region are weighted in the average by the number of 
pixels in each subraster. There are usually 4 dark measurements available. 
If any of these individual readings deviates strongly (by more than 
3$\sigma$) from the mean (likely due to a particularly energetic cosmic 
ray hit), then that value is discarded.

\subsubsection{Flatfield Corrections\label{flat}}

The MOST instrument was designed to obtain fixed extended pupil images of 
a single bright target star through a Fabry microlens. In this way, the 
Fabry Imaging measurements are insensitive to pointing wander of the 
satellite.  The Direct Images, however, do reflect the pointing errors.
In data from the satellite commissioning and its early scientific operations,
pointing errors of up to about 10 arcsec led to wander of the Direct Images
of up to 3 pixels on the CCD.  This made precision photometry vulnerable to
uncalibrated sensitivity variations among adjacent pixels.  The pointing 
performance of the satellite has been improved dramatically since its early
operation, now consistently giving positional errors of about $\pm1$ arcsec
$\sim$ 0.3 pixel $rms$ (see \S\ref{cent}).

Despite the lack of an on-board calibration lamp, it is possible to recover
some flatfielding information for each subraster by exploiting intervals of
high stray light during certain orbital phases.  Frames containing no 
detectable stars are chosen from the data set.  Such frames can be
obtained when the satellite is commanded to point to an ``empty''
field (with only stars much fainter than V = 13), or occasionally,
when the satellite loses fine pointing and the stars wander outside their 
respective subrasters.  During two weeks of observing, over which 50,000 - 
100,000 individual exposures are typically obtained, about 2000 - 4000 
flatfielding frames are usually available from when the satellite
loses fine pointing.

The stray light can produce a strong spatial gradient across the CCD, so 
a 2-dimensional polynomial is fitted to all the available flatfielding 
pixels and removed. For each subraster, the mean pixel value is measured 
and compared to each individual pixel value; a linear fit to the correlation
is made.  The slope of this relationship is the relative gain and the zero 
point is the dark current. The relative pixel gains are found to vary by 
less than 2\% across any individual subraster, with the standard
deviation for any individual pixel less than 0.5\%.

\subsubsection{Star Detection and Centroiding}\label{cent}

We employ two methods for star identification. The first method involves
deconvolving the individual subrasters with a model Point-Spread Function 
(PSF). The model PSF is created by registering subrasters containing stars 
and mapping the pixel values onto an oversampled grid to create a model 
profile with twice the resolution of a real image. After deconvolution, 
the strongest source is chosen as the target star for photometry. The 
second method starts by prompting the user to identify by eye stars on 
each subraster via a graphical interface, which is particularly useful for
fields containing several stars.

Once the stars have been identified, centroids are computed by an 
intensity-weighted average on a 5x5 grid around each object of interest. 
On the first frame, the centroids for all the stars are saved as a master 
grid used to check the positioning for all subsequent frames. After the 
first frame the strongest source is always selected using the deconvolution 
technique and its position is calculated with intensity-weighted means. 
This is done for each subraster and the offsets are compared to the master 
grid of centroids.  If the position of a star varies by more than 2 pixels, 
then that centroid is corrected by using the average offset from the 
remaining objects. This correction is particularly important when cosmic 
rays interact with the detector and can mimic a stellar PSF, leading to 
the occasional spurious star identification.

\begin{figure}
\epsscale{0.50}
\begin{center}
\plotone{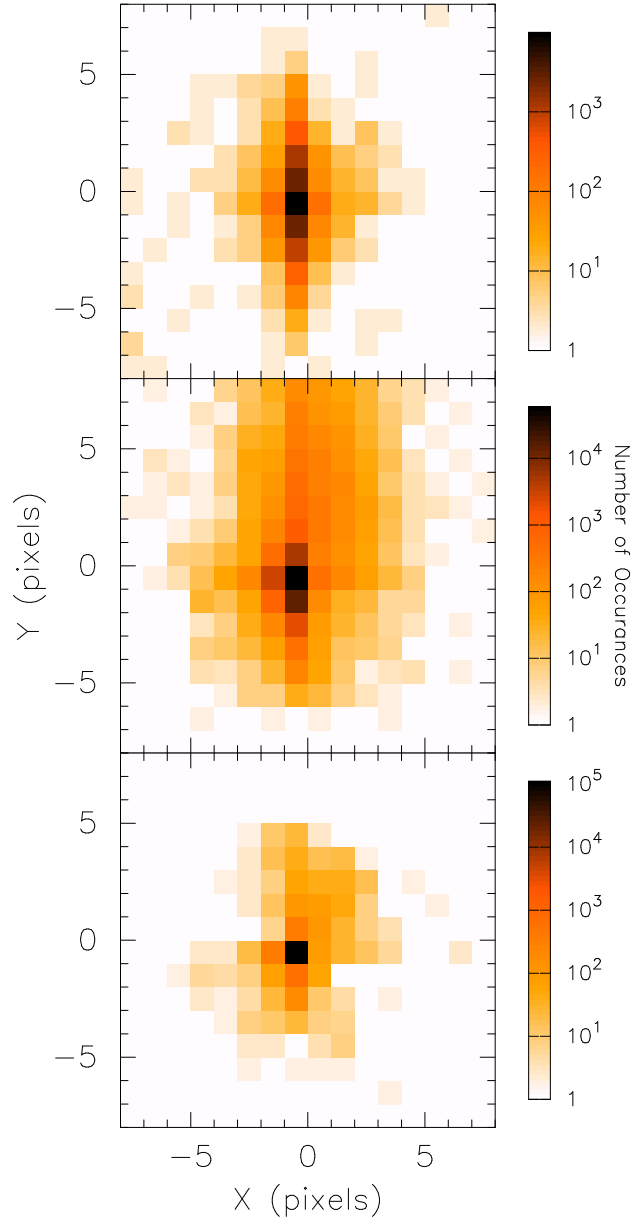}
\end{center}
\caption{The pointing performance for three different targets.  HD 263551
is shown in the top panel, HD 61199 in the middle panel and HD 209458 in
the bottom panel. The top panel represents the typical point
performance in early satellite operations. There is a substantial drift
in the Y-direction.  In the middle panel, the Y-axis drift has been
largely eliminated.  The bottom panel shows the current and much
improved pointing performance.}
\label{fig:positions}
\end{figure}

The pointing performance of MOST has been dramatically improved since its
original on-orbit commissioning, thanks to upgraded software and a better 
understanding of the mechanical performance of the reaction wheels in its
attitude control system. In Figure \ref{fig:positions}, the
distribution of pointing centroids for three fields observed from
December 2003 to August 2004 are shown.  The first is for the star HD
263551 in the field of $\xi$ Geminorum (observed during 20 - 28 December
2003); the second, for HD 61199 in the field of Primary Target Procyon
(8 January - 9 February 2004); and the third, for HD 209458 itself
(14 - 30 August 2004). The standard deviation for each target in x and
y pixel directions  
are (0.985,1.945), (0.947,1.837), and (0.309,0.403), respectively.  The 
substantial improvement in tracking stability has resulted in improved 
photometry precision as flat fielding errors have become less relevant.

\subsubsection{Stray Light and Background Determination\label{gradient}} 

In order to remove the strong background gradients associated with stray 
light, a 2-dimensional 2nd order polynomial is fitted to the
subrasters based on the  
pixel co-ordinates on the CCD and subtracted.  A {\it sky} radius is 
defined, centred on selected stars.  Only pixels outside this radius are 
included in the polynomial fit, to minimize influence from stellar sources.  

Once the gradient has been suppressed, the background level for each 
subraster is determined by rejecting pixels with levels more than 2.5 
standard deviations from the median to eliminate cosmic ray hits.  The
rejection of pixels is iterated  
until the median converges, and the background is defined as the mean of
the remaining pixel set.  The median is chosen for the calculation of
the standard deviation since the small subraster sizes limit the total 
number of pixels available for background determination, making the mean 
sensitive to errant values, mostly due to cosmic ray events.

\subsubsection{PSF Fitting and Adding Up the Starlight}\label{psf}

After determination of the star positions and the background, the PSF for
each star is fitted by either a Moffat profile \citep{mof69} or a Gaussian 
function.  In general it is found that the Moffat profile provides a better 
approximation of the PSF shape (even though it was designed to reproduce
stellar images smeared by atmospheric seeing).  The difference between
a Gaussian and Moffat profile is that the wings of the latter fall off
much more slowly accounting for the scattering profile at large
off-axis distances, whether in the Earth's atmosphere or in space,
including telescope optics.  The PSF is computed 
independently for each subraster using the Levenberg-Marqardt approach 
\citep{pre92} to find the best fit parameters.  The background level can be 
allowed to vary with the fit minimization procedure, but we find that 
better photometry is obtained by fixing the level as determined by the 
procedure outlined in \S\ref{gradient}.

For data sets obtained early in the mission lifetime, when the tracking
performance was not ideal, such as with $\xi$ Gemini, multiple images
of the same source can appear on the subraster.  In these cases, once
the initial PSF fit has been made, the fit is removed and the residual
image is deconvolved using a predetermined instrumental PSF.  The
strongest peak is then identified and the original subraster data is
fitted with both centroids.  This process is iterated until the
deconvolved image has no significant peaks.

Once the stellar source has been modeled, the total flux is estimated
by using aperture photometry for the centre of the model fit and using
the model fit for the faint extended wings.  For cases such as $\xi$
Gemini, this step is important as the aperture accounts for the
smearing effects of pointing jitter not included in the model and the
model allows an estimation of stellar flux located 
outside the subraster.  The FWHM of the PSF under good tracking is
found to be about 2 pixels, but the stellar source can be easily
traced out to a radius of 8 pixels, meaning for large pointing
deviations a significant portion of flux can lie outside the
subraster.  With HD 209458 where the pointing accuracy is
greatly improved, only the PSF model is necessary to determine the
brightness of the source.  The final magnitude is defined as
\begin{equation}
mag = 25.0 - 2.5*{\rm log}\left(\frac{F_p + F_a}{E_t} g\right)
\end{equation}
as the standard conversion between magnitude and flux,
where $F_p$ is the flux in ADU (Analogue-to-Digital 
Units) measured from the PSF fit, $F_a$ is the
flux residual inside a small aperture in ADU, $E_t$ is the exposure
time in seconds and $g$ is the gain in $e^-$/ADU.  The zero point of
25 has been arbitrarily chosen.                  

\subsubsection{Removal of Stray Light Effects}

\begin{figure}
\begin{center}
\plotone{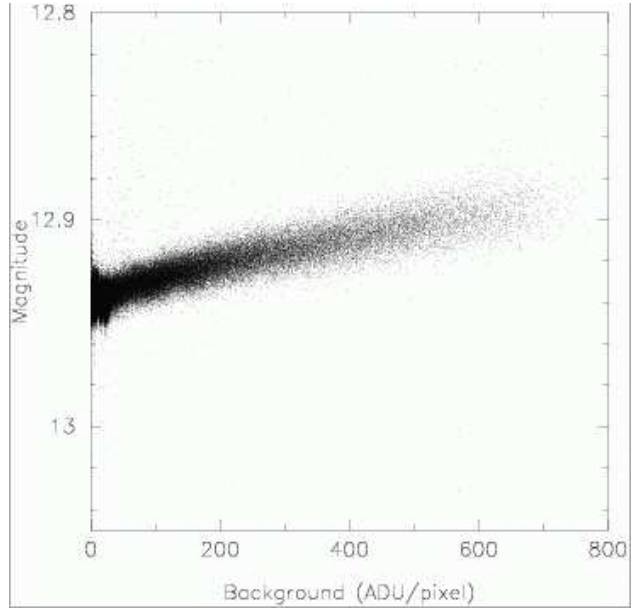}
\end{center}
\caption{The relationship between the instrumental magnitude and the
  background level as measured on the CCD frame.}
\label{fig:skymag}
\end{figure}

Once the instrumental photometry has been extracted, variable stray 
light effects must be removed.  It was discovered that the background 
as determined in \S\ref{gradient} level needs to be scaled to properly 
remove the contribution from stray light.  The cause of this effect
may because the 2D fits (see \S\ref{gradient}) are not sufficient.
With the small number of subraster pixels a higher order polynomial
is not stable enough, hence the estimation of the background level is
not optimum.  Regardless, it can be corrected.  In Figure
\ref{fig:skymag}, the  relationship between instrumental magnitude and
the background level is  shown.  A fit is made between the stellar
flux and the background level  either with a polynomial or a cubic
spline depending on the complexity  of the relationship.  If the
relationship is modeled with a cubic  
spline, then the original data are binned before the fit, with 500 
data points per bin and a minimum bin width of 10 ADU per pixel.  In
most cases the peak-to-peak amplitude of stray light  
variations is reduced by this approach to a few $\times$ 0.1 millimag. It
can be further reduced by techniques such as subtracting a running, 
averaged background phased to the orbital period \citep{ruc04}.  The 
advantage of the simple approach taken here is that no prior knowledge of 
the orbital period is required and the amplitude of the stray light 
component is allowed to vary from orbit to orbit with changes in the 
Earth's albedo.

\section{HD 209458}

\subsection{Observations}

Observations of HD 209458 were made during 14 - 30 August 2004, as
part of a relatively short trial run on this star.  This star is
accessible to MOST in its Continuous View Zone (see \S\ref{intro}) for
approximately 45 days, but this was the first opportunity to test the
Direct Imaging mode on an exoplanetary system target.  The exposure
time was 1.50 s (regulated to better than 1 $\mu$s).  Sampling rates
were varied.  During most of the run, to stay within downlink margins
for the MOST ground station network, data were collected at a rate of
5 per minute.  For 15 hours centred on the predicted times of
exoplanetary eclipse, the sampling rate was increased to 10 per
minute.  This improved our sensitivity to eclipses, while staying within
the MOST on-board buffer capacity so no data would be lost in the
event of a malfunction of one of the three ground stations.

There were data losses in the first third of the run due to crashes of
the satellite's control system (ACS) when subtle bugs in new software
(which had operated smoothly during the previous month of observations
on another target) manifested themselves.  Once the problem was
traced, the previous version of the software was uploaded to the
satellite and observations continued with only one brief interruption
due to a cosmic-ray-induced crash.  The overall duty cycle of the raw
photometry is about 85\%.

Near the solstices, the contributions of scattered Earthshine (stray
light) in the MOST focal plane reach their peak.  For a bright Fabry
Imaging target, the stray light can be less than 1\% of the stellar
signal.  for a fainter Direct Imaging Target like HD 209458 (V $\sim$
7.7), there is an interval during each orbit where the stray light is
high enough to seriously degrade the quality of the photometry.
Whenever the background level exceeded 300 ADU/pixel in an exposure,
that measurement was excluded from the analysis.  This represents
about 25 minutes from each 101.413-min orbit, reducing the
total duty cycle to 53\% but still providing thorough coverage of the times of
transit and eclipse.

There is an unfortunate coincidence between the orbital period of the HD
209458 exoplanet around its parent star and the orbital period of the
MOST satellite around the Earth.  The ratio of the two periods is
50.049, so that over the 2-week span of the MOST observations, the
satellite orbital period maintains rough phasing with the exoplanet
epheremis.

MOST was designed to be a non-differential photometer in its Fabry
mode, but in the Direct Imaging, there can be other stars in the field
appropriate as photometric comparisons.  In the HD 209458 field, two
other relatively bright stars were also monitored:  HD 209346 (V = 8.33)
and BD+18 4914 (V = 10.60); see Table \ref{ta:target}.  Even the
brighter of these two is almost 0.7 mag fainter than HD 209458, so
differential photometry tends to add noise to the point-to-point
scatter and degrade the sensitivity to eclipses.  However, it does
provide a check for slow instrumental and/or environmental drifts.
It was noticed that the data for HD 209458 during JD 1693 - 1693.5
(JD - 2451545.5) showed a slow trend that was not present in HD
209346.  This change was approximately 0.005 mag.
This could represent intrinsic variability in the star HD 209458a.
Since we are interested in variations due to phase changes
of the planet over the orbital period one cannot fit the trend and
restore the data without perturbing trends due to the planet so it was
excised from this analysis. This stretch of data did not overlap with
a phase of transit or eclipse thus its exclusion has minimal effect
for detection of the secondary eclipse. 

\begin{figure}
\begin{center}
\plotone{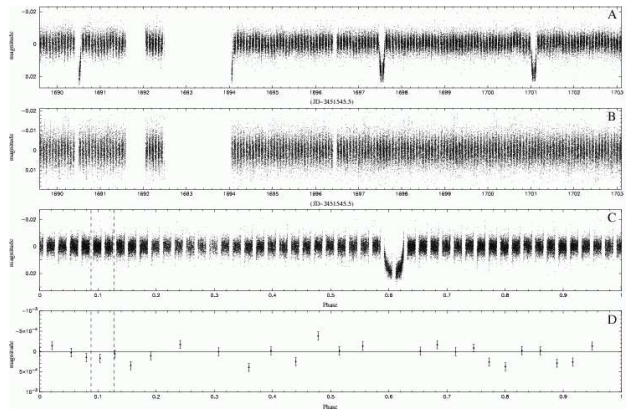}
\end{center}
\caption{Panel A shows the light curve for HD 209458,
  panel B shows HD 209346, panel C shows data for HD 209458 phased
  with the planet orbital period and panel D shows the data phased
  with the planet orbital period and heavily binned.  The dashed lines
  in panels A and B mark when eclipse occurs. }
\label{fig:data}
\end{figure}

The reduced light curves for HD 209458 and HD 209346 are shown in 
Figures \ref{fig:data}a and \ref{fig:data}b. The primary transits in the HD
209458 system are obvious, as are the regular gaps during each satellite
orbit in which data were removed due to high stray light.

Next any residuals due to the orbital period where removed from both
light curves by fitting an equation of the form
\begin{equation}\label{fourier}
mag = A_0 + \sum _{j=1,n} A_j \cos (j\omega t + \phi _j),
\end{equation}
where $A_0$ is the mean magnitude, $A_j$ and $\phi _j$ are the
amplitude and phase coefficients for the cosine series.  $\omega$ is
($2\pi$/period) and n=8 was chosen to accomidate the non-sinusoidal
shape of the stray light residuals.  Long term trends in the 
light curve for HD 209458 were removed by binning the comparison star
with 0.2d bins and interpolating with a cubic spline. The last step
was to filter 
a 1-cycle/day variation in the data (peak-to-peak amplitude ~ 1.3
mmag) in the data likely due to the fact that the MOST satellite
returns to a similar albedo feature on the Earth each day.

The final reduced data are presented in a phase diagram in Figure
\ref{fig:data}c, using the ephemeris for HD 209458b determined by
\citet{wit05}, with an exoplanetary period of 3.52474554 d.  The
"tire-track" pattern in the data 
is due to the intervals of high stray light subtracted from each MOST
orbit and the near-harmonic relationship between the orbital periods
of the exoplanet and the MOST satellite.  Unfortunately, intervals of
high stray light happened in this data set to coincide with the phases
of ingress, minimum and egress of the exoplanetary transits.  The
primary transit is obvious in Figure \ref{fig:data}c; the range of
phases of the secondary eclipse is marked in the figure.

\section{Upper Limit on the Eclipse}
\label{sec-uleclipse}
\begin{figure}
\begin{center}
\plotone{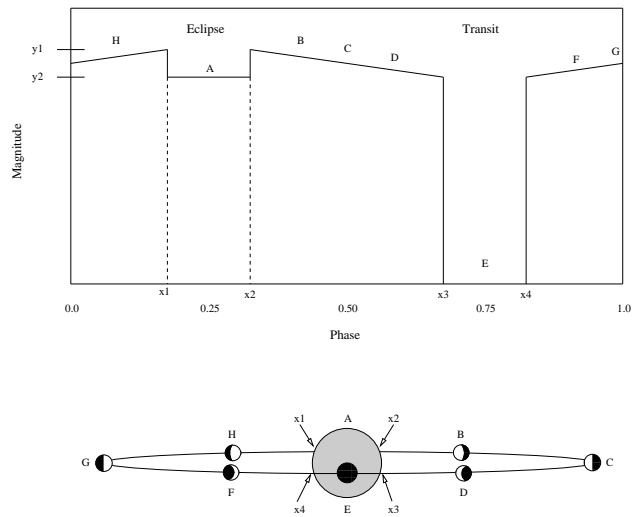}
\end{center}
\caption{The eclipse model schematic. Point A marks when the eclipse is occurring
and the magnitude at this point is defined at $y_2$. The points on
ingress and egress are labeled as $x_1$ and $x_2$. As the planet
moves along its orbital path the light curve proceeds through points
B, C and D.  When the transit occurs the total flux from the system
drops and is marked as Point E.  The start and finish of the transit
is labeled as $x_3$ and $x_4$.  The planet and light curve then
proceeds through points F, G and H towards the occurrence of another
transit.}
\label{fig:model}
\end{figure}

To search for evidence of an eclipse in the light curve, we apply a 
model of the light variations due to the reflected light of the exoplanet
which is shown schematically in Figure \ref{fig:model}.  We neglect
any intrinsic  
variability of the star in this version of the model.  During segment A of 
the phase diagram, the planet is totally eclipsed by the star, and the 
system has a total magnitude of $y_2$.  As the planet moves in its orbit 
towards inferior conjunction at point D (middle of transit), the illuminated
fraction of the planet disk decreases. The points of maximum brightness
occur at phases $x_1$ and $x_2$ where the planet is almost fully illuminated;
minimum brightness (outside of transit) occurs at phases $x_3$ and $x_4$, 
where we see the night side of the planet.

We can set up a simple model to approximate the expected phased light curve 
by a series of straight lines connecting the points $x_1, x_2, x_3$
and $x_4$.  In reality these points should be connected by slowly
varying curves, but our first order approximation is valid considering
the star-planet flux ratio is greater than 10 000.

Thus if $x_1 \leq x < x_2$ we have
\begin{equation}\label{model1}
f(x;y_1,y_2) = y_2;
\end{equation}
if $x_2 \leq x < x_3$
\begin{equation}\label{model2}
f(x;y_1,y_2) = \left(\frac{y_2-y_1}{x_3-x_2}\right)(x-x_2)+y_1;
\end{equation}
and if $x_4 \leq x < x_1$
\begin{equation}\label{model3}
f(x;y_1,y_2) = \left(\frac{y_1-y_2}{x_1-x_4}\right)(x-x_4)+y_2.
\end{equation}
Since, we know a priori when are the phases of eclipse and transit, we can 
define $x_1$=0.087, $x_2$=0.127, $x_3$=0.587, and $x_4$=0.627.  These
were chosen to bracket the duration of the eclipse and transit and
also reside in 
the gaps of the light curve where data was cut due to high levels of
Earthshine with a duration of 3.4 hours ($x_2$ - $x_1$).  The depth of
the eclipse will be, by definition, the difference between $y_1$ and $y_2$.

\begin{figure}
\begin{center}
\plotone{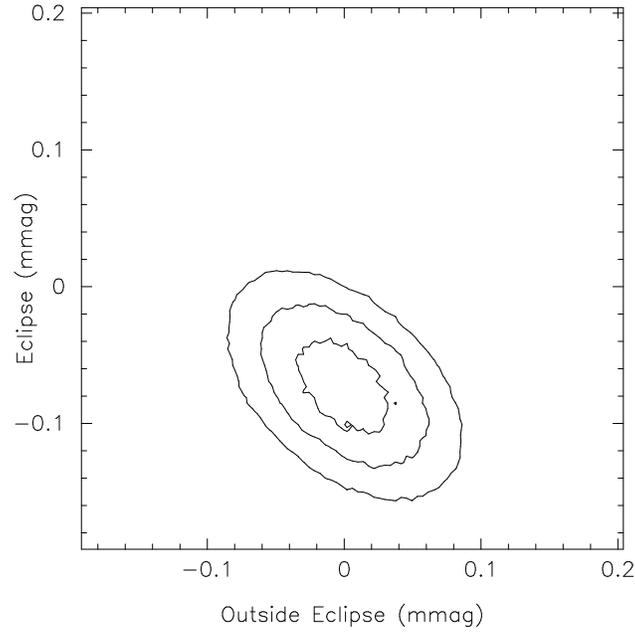}
\end{center}
\caption{Confidence levels for eclipse using all data. The contours
  represent 68.3, 95.4 and 99.7\% confidence levels.}
\label{fig:confidenceall}
\end{figure}

\begin{figure}
\begin{center}
\plotone{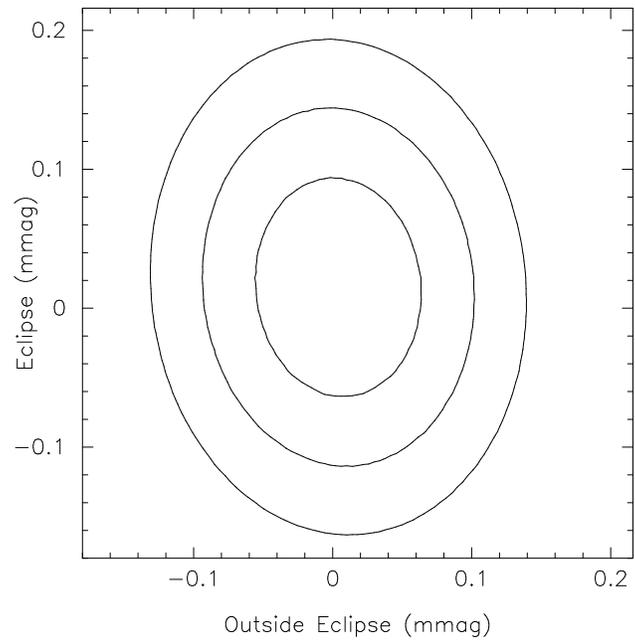}
\end{center}
\caption{Confidence levels for eclipse using only data around the
  occurrence of the eclipse. The contours represent 68.3, 95.4 and
  99.7\% confidence levels. }
\label{fig:confidence}
\end{figure}

We choose a Bayesian approach to directly compute the probability
distribution of the eclipse depth as we descripe below.
To find the most probable values for $y_1$ and $y_2$ based on the data, we 
start with Bayes' Theorem  
\begin{equation}
p(H|D,I)=\frac{p(H|I) p(D|H,I)}{p(D|I)},
\end{equation}
where we have adopted the formalism of \citet{gre05}.  $H$ is our hypothesis 
of interest, $D$ is our data set, and $I$, our current state of knowledge.  
Starting with $p(y_1,y_2|D,I)$, we obtain an expression for the probability 
of $y_2$ by marginalizing over $y_1$, giving us
\begin{equation}\label{eq:marg}
p(y_2|D,I)=\frac{p(y_2|I)}{p(D,I)} \int p(y_1|I) p(D|y_1,y_2,I) dy_1,
\end{equation}
where we have chosen a Jeffrey's Prior
\begin{equation}
p(y|I)=\frac{1}{y\ ln(\frac{y_{max}}{y_{min}})}
\end{equation}
to give equal probability to all parameters where $y_{min}$=-0.3 mmag and 
$y_{max}$=0.3 for both $y_1$ and $y_2$.  This where chosen based on
the scatter seen in the binned phased light curve shown in Figure 4d.
We repeated all calculations using a uniform prior and  
found no change in our results as our parameter range is well constrained by 
the data set.  Since the dataset is well constrained it should be
noted that a maximum likelihood analysis will give similar results as
the probability distribution is not strongly affected by our prior. 
The likelihood function is given by  
\begin{eqnarray}
p(D|y_1,y_2,I) & = & {\rm Exp}[-\chi^2]\
\prod_{i=1}^n\frac{1}{\sqrt{2\pi}\sigma_i},\\ 
\chi^2          & = & \sum_{i=1}^n \frac{(d_i -
  f(x_i;y_1,y_2))^2}{\sigma_i^2},
\end{eqnarray} 
where $f(x;y_1,y_2)$ is given by Equations \ref{model1}-\ref{model3},
, $d_i$ are the observations, $\sigma_i$ are the photometric errors for
each observation, n is the number of observations and the 
normalization factor is
\begin{equation}
p(D|I)=\int \!\!\! \int p(y_1|I) p(y_2|I) p(D|y_1,y_2,I) dy_1 dy_2.
\end{equation}
Using all available data except during high levels of Earthshine we
find that ($y_2 - y_1$) is -0.08 $\pm$0.05 mmag
(90\% confidence), which would imply the system is getting brighter
during the planetary eclipse.  The confidence levels are plotted in
Figure \ref{fig:confidenceall}.  In this analysis, we have not accounted 
for potential intrinsic variations in the star.  To make our analysis 
less sensitive to longer-term variability in the star, we restricted
the data set to phases bracketing the eclipse, defining three bins of
equal size: one extending from $x_1$ to $x_2$ and two other bins of 
equal size ($x_2-x_1$) adjacent to the eclipse.  Our initial choice of
$x_1$ and $x_2$ to occur in the data gaps also means that the new
adjacent boundaries occur at data gaps.  This means that each
bin will have approximately the same signal-to-noise ratio.  We then
recalculated  the probabilities for $y_1$ and $y_2$, which we plot in Figure
\ref{fig:confidence} using 1, 2 and 3 $\sigma$ confidence levels for a
2-dimensional distribution. The best fit parameter for the value of
$y_2-y_1$ are then obtained from maximum of equation \ref{eq:marg}
where we have adjusted the mean of the dataset so that $y_1$ is equal
to zero.  Our detection limits are 
summarized in Table \ref{ta:data}.  Our fit is consistent with no
detection of an eclipse, but it does allow  
us to place an upper limit and confidence interval on the photometric 
depth of the eclipse.  We also repeated our analysis via a
chi-squared minimization analysis.  In this case one minimizes
Equation 10 and we get a best fit for $y1-y2$ of 0.018 mmag.  Thus our
answers remain unchanged under this type of analysis 
as our choice of prior does not strongly affect our probability
distribution.

\section{Upper Limit on the Albedo of HD 209458b}

In this section we derive the upper limit on the 
planet-star flux ratio and convert it to an upper limit on the 
geometric albedo. 


We convert the secondary eclipse upper limit value from  magnitudes to a
planet-star flux ratio using the standard equation for the definition
of magnitude,
\begin{equation}\label{eq:flux}
y_1 - y_2 = -2.5 {\rm log_{10}} \frac{F_p + F_*}{F_*}.
\end{equation}
We take the error on the secondary eclipse $y_1-y_2$ as the eclipse
upper limit, because the scatter in the data is larger than the
putative eclipse measurement. Table~2 lists the upper limit (derived
in \S\ref{sec-uleclipse}) for different confidence levels.  Here we
work with the 1~$\sigma$ (or 68.3\% confidence level) eclipse upper
limit value, which is 0.053 mmag.

Using the eclipse upper limit $y_1-y_2 = 0.053$~mmag, 
the planet-star flux ratio upper limit is 
\begin{equation}
\frac{F_p}{F_*} \leq 4.88 \times 10^{-5}.
\end{equation}
The albedo is a more intuitive form of the planet-star flux ratio,
since it describes the fraction of stellar radiation that is
scattered by the planet.

The geometric albedo, $A_g$ is the quantity relevant to the MOST
measurement. $A_g$ is defined as the ratio of the planet's luminosity
at full phase to the luminosity from a Lambert disk\footnote{A
Lambertian surface is an ideal, isotropic reflector at all
wavelengths.} with the same cross-sectional area as the planet.  $A_g$
is equivalent to the fraction of incident stellar radiation scattered
in the direction of the observer for planetary full phase (if the
stellar intensity is spatially uniform).  The MOST measurement occurs
over a range covering 7 degrees on either side of full phase, and is
close enough to full phase that we use the geometric albedo
terminology\footnote{Strictly speaking $A_g$ is defined only at full phase.
However, the MOST measurements are close enough to full phase
that we use this familiar term.}. $A_g$ is usually specified at a particular
wavelength, so we use $A_{g {\rm MOST}}$ for the geometric albedo
integrated over the MOST bandpass (400 to 700 nm, see Table~3) and
$A_{g {\rm TOT}}$ for the geometric albedo integrated over all
wavelengths.

$A_g$ is related to the planet-star flux ratio by
\begin{equation}\label{eq:ag}
\frac{F_p}{F_*} = A_g \left(\frac{R_p}{a}\right)^2,
\end{equation}
where $R_p$ is the planetary radius and $a$ is the semi-major axis.
Taking $R_p = 1.35\pm 0.06 R_J$ (Brown et al. 2001) and $a = 0.046$~AU
(Mazeh et al. 2000), $A_{g {\rm MOST}} \leq 0.25$. The 5\% uncertainty
in planet radius translates into a $\sim$~10\% uncertainty in $A_g$.
We therefore more accurately state our HD 209458b 1$\sigma$ geometric
albedo upper limit as
\begin{equation}
A_{g {\rm MOST}} \leq 0.25 \left (\frac{1.35}{R_p}\right)^2.
\end{equation}

\section{Discussion}

HD 209458b's geometric albedo of $\leq 0.25$ is a relatively low
value. The solar system giant planet albedos in the MOST bandpass are
$ \gtrsim 0.4$, with Jupiter's value 0.5 (computed from data in 
\citet{kar98}\footnote{The Karkoschka albedos are measured at 6.8,
5.7, 0.7, and 0.3 degrees away from full phase for Jupiter, Saturn,
Uranus, and Neptune respectively. The albedos have an uncertainty of
4\%. Jupiter's and Saturn's albedo are probably about 5\% higher at
full phase where the definition of geometric albedo formally applies
\citep{kar98}.}; see Table~4). HD 209458b is therefore less than half
as bright as Jupiter at the MOST wavelengths.

The solar system giant planets all have bright cloud decks (water ice
or ammonia ice) which cause them to be bright in the MOST bandpass.
HD 209458b is an order of magnitude hotter than Jupiter, far too hot
for water or ammonia clouds to be present. HD 209458b may have clouds
in its atmosphere, but composed of high-temperature condensates such
as silicates or solid iron, instead of from ices.  Clouds at high
altitude are consistent with previous observations of the HD209458b
atmosphere including: a primary transit low sodium absorption
\citep{cha02}; a primary transit CO non-detection \citep{dem05b}; and
a secondary eclipse non-detection of H$_2$O at 2.2~$\mu$m
\citep{ric03}. If clouds are present in the HD 209458b atmosphere, the
low $A_{g {\rm MOST}}$ rules out any bright clouds at high
altitudes.  Unlike ice clouds, high temperature-condensate clouds may 
be dark if they consist of small particles or are predominantly Fe
(see Figure~5 in \citet{sea00}). If HD 209458b does not have clouds,
strong sodium and potassium atomic absorption could be present on the
day side and cause a low albedo in the MOST bandpass. While $A_{g {\rm
MOST}}$ is not definitive, it is a key constraint on HD 209458b
atmosphere models with specific regard to the thickness, altitude,
composition, and particle size distribution of clouds.  Our
measurements are consistant with the results of \citet{col02} and 
\citet{lei03} that also find low albedo upper limits for the short
period planetary companions of $\upsilon$ And and HD 75289 which have
orbital periods of 4.6d and 3.5d, similar to HD 209458b.

With an upper limit determined for the measured geometric albedo a
natural question is can we estimate 
the Bond albedo?  The Bond albedo, $A_B$, is the total radiation
reflected from the planet compared to the total incident radiation,
i.e. the total amount of radiation reflected in all directions
integrated over all wavelengths. $A_B$ is an important physical
parameter because it specifies the amount of stellar radiation
absorbed by the planet and hence the equilibrium temperature of the
planet,
\begin{equation}
\label{eq:Teq}
T_{eq} = T_{*} \left( \frac{R_*}{2 a} \right)^{1/2} \left[f (1 -
A_B) \right] ^{1/4},
\end{equation}
where $R_*$ is the stellar radius, $T_*$ is the stellar effective
temperature, $a$ the semi-major axis, and $A_B$ the Bond albedo.  Here
$f$ is the proxy for atmospheric circulation, $f=1$ if the absorbed
stellar radiation is redistributed evenly throughout the planet's
atmosphere (e.g., due to strong winds rapidly redistributing the heat)
and $f=2$ if only the heated day side reradiates the energy.
Figure~\ref{fig:teq} shows $T_{eq}$ for the HD209458 parameters
\citep[$T_* = 6000$~K, $R_*=1.18R_{\odot}$, and
$a=0.046$~AU;][]{maz00, cod02}.  The upper left corner represents our
parameter range space based on 1 $\sigma$ limits.

In principle $A_B$ could be measured for a transiting extrasolar
planet if its brightness at all phases could be measured in a
wavelength range that encompasses all the planet's scattered
light. However, HD 209458 is too faint for such a measurement by MOST,
and MOST's bandpass has a cutoff at 0.7 microns (Table
3). Nevertheless, we estimate an $A_B$ upper limit for HD209458b,
based on the solar system planet albedos, the $A_{g {\rm TOT}}$/$A_B$
relation, and model atmosphere considerations.

The solar system giant planets all have $A_B > A_{g {\rm TOT}}$, as
illustrated in Figure 8. This can be understood by considering a
Lambertian planet, with $A_B = 1.$ $A_{g {\rm TOT}}$ would have to be
less than one, since $A_g$ includes only the radiation scattered back
toward the observer. More precisely, for a Lambert sphere $A_B =
1.5~A_{g {\rm TOT}}$ (Lambert's law is the dotted line in
Figure~8). Under the reasonable assumption that HD 209458b is a gas
giant planet with a thick atmosphere and no reflecting surface, we can
confine our attention to the very general theoretical case described
by a semi-infinite atmosphere model. In this case the physically
relevant albedo parameter space in Figure 8 is bound by the dashed and
dotted lines, $0.67 < A_{g {\rm TOT}}/A_B < 1$ \citep[e.g.,][]{sob75}.
Indeed, the solar system giant planet albedos comply
\citep[][reproduced in Table 4]{con89}.  Therefore, under the simplest
case assumptions about the atmosphere of HD 209458b, we can use the
isotropic scattering limit (dotted line) in order to derive an upper
limit on its $A_B$.  Multi-layered atmosphere models and other
complications can produce geometric albedos below that limit by no
more than $\sim$10\% (see \citet{sob75}, Ch.9).

One further assumption is required in order to estimate HD 209458b's
$A_B$ from the $A_{g {\rm MOST}}$ upper limit: $A_{g {\rm TOT}} = A_{g
{\rm MOST}}$.  We first note that the $A_{g {\rm TOT}} < A_{g {\rm
MOST}}$ for solar system planets because of strong CH$_4$ absorption
redward of the MOST bandpass, but blueward to the wavelength where
their thermal emission dominates over scattered radiation.  HD 209458b
is an order of magnitude hotter than the solar system giant planets
and should differ.  If HD209458b were a blackbody emitter, its thermal
emission would peak around 2--5 microns (depending on its actual
equilibrium temperature).  HD209348b, however, is expected to deviate
significantly from a blackbody. Near-IR molecular absorption could
induce thermal emission as short a wavelength as 0.8 or 0.9 microns
(see \citet{sea05} and references therein).  In low-geometric-albedo
models, the thermal emission could dominate over scattered radiation
at such short wavelength and $A_{g {\rm TOT}} \simeq A_{g {\rm MOST}}$
is not too unreasonable.  With $A_{g {\rm TOT}}$ and the $0.67 < A_{g
{\rm TOT}}/A_B < 1$ argument (Figure~8), we estimate for HD 209458b
that $A_B \lesssim 0.375$.

 From equation~(\ref{eq:Teq}) this value
of $A_B$ gives $T_{eq} > 1300$~K. In Figure~\ref{fig:teq} we show how
this estimated $A_B$ value together with the Spitzer/MIPS 24 $\mu$m
brightness temperature measurement of 1130~K \citep{dem05b} constrain
the overall $T_{eq}$ of HD209458b.

A more robust $A_B$ estimate requires detailed model computations,
beyond the scope of this paper.  Because of a wide viable parameter
space \citep[][and references therein]{sea05}, however, more data are required
before a definitive HD209458b model atmosphere can be
computed.  Indeed strong H$_2$O near-IR absorption has not yet been
detected \citep{ric03, sea05}.  Upcoming data, including Spitzer
programs for secondary eclipse thermal emission measurements
(photometry from 3.5 to 8 microns and spectra from 7.4 to 14.5 microns)
and HST primary transit data analysis for H$_2$O absorption (T. Brown
2005, private communication) will help constrain models and thus
the $A_B$ estimate.

In summary, MOST has observed HD209458 for 14 days and 
we have derived an
upper limit on the 
planet-star flux ratio of $4.88 \times 10^{-5}$, corresponding to a
geometric albedo of $\leq 0.25$ at the $1 \sigma$ level. These numbers
at the 3$\sigma$ level are $1.34 \times 10^{-4}$ and 0.68
respectively.  During a second HD209458b observing campaign three
times longer than the one described in this paper, MOST will either
detect the secondary eclipse or put a significant limit on it to a
geometric albedo of 0.13 (1 $\sigma$) or 0.34 (3 $\sigma$).  As the
only existing constraint on scattered light from HD209458b, the MOST
geometric albedo upper limit will play a pivotal role in constraining
HD209458b atmosphere interpretations.

\acknowledgments

The contributions of JMM, DBG, AFJM, SR, and GAHW are supported by
funding from the Natural Sciences and Engineering Research Council
(NSERC) Canada.  RK is funded by the Canadian Space Agency. WWW
received financial support from the Austrian Science Promotion 
Agency (FFG - MOST) and the Austrian Science Fonds (FWF - P17580)

\onecolumn{

\begin{deluxetable}{lccccc}
\tablecolumns{12}
\tablewidth{0pc}
\tablecaption{Coordinates of Direct Imaging Target Stars.}\label{ta:target}
\tablehead{
\colhead{ID} & \colhead{R.A.} & \colhead{Dec.} &
\colhead{Mag. (V)} & \colhead{(B-V)} & \colhead{Spec. Type}
}
\startdata
HD 209458 & $22^{{\rm h}}03^{{\rm m}}10.80^{{\rm s}}$ &
$+18\degr 53' 04.0''$ & 7.65 & 0.53 & G0V\\
HD 209346 & $22^{{\rm h}}02^{{\rm m}}21.33^{{\rm s}}$ &
$+18\degr 49' 59.2''$ & 8.33 & 0.25 & A2\\
BD+18 4914 & $22^{{\rm h}}02^{{\rm m}}37.70^{{\rm s}}$ &
$+18\degr 54' 02.6''$ & 10.6 & 0.5 & F5 \\
\enddata
\end{deluxetable}

\begin{deluxetable}{lcccc}
\tablecolumns{5}
\tablewidth{0pc}
\tablecaption{Eclipse of HD 209458{\rm b}.}\label{ta:data}
\tablehead{
\colhead{} & \colhead{} & \multicolumn{3}{c}{Confidence Level (\%)} \\
\cline{3-5}\\
\colhead{Parameter} & \colhead{Best Fit} & \colhead{68.3} &
\colhead{95.4} & \colhead{99.7}
}
\startdata
$y_2 - y_1$ (mmag)& 0.013 & 0.053 & 0.105 & 0.145 \\
$F_p/F_*$ & 1.20 $\times$ $10^{-5}$ & 4.88 $\times$ $10^{-5}$ &
9.67 $\times$ $10^{-5}$ & 1.34 $\times$ $10^{-4}$ \\ 
$A_g$ & 0.06 & 0.25 & 0.49 & 0.68 \\
\enddata
\end{deluxetable} 

\begin{deluxetable}{cccccc}
\tablecolumns{6}
\tablewidth{0pc}
\tablecaption{MOST bandpass filter characteristics.}\label{ta:filter}
\tablehead{
\colhead{Wavelength} & \colhead{Transmission} & \colhead{Wavelength} & 
\colhead{Transmission} & \colhead{Wavelength} & \colhead{Transmission}\\
\colhead{(nm)} & \colhead{(\%)} & \colhead{(nm)} & \colhead{(\%)} &
\colhead{(nm)} & \colhead{(\%)} 
}
\startdata
400 &  0.00 & 530 & 83.40 & 660 & 79.20 \\
410 & 19.80 & 540 & 74.90 & 670 & 82.40 \\
420 & 58.20 & 550 & 82.20 & 680 & 78.10 \\
430 & 62.50 & 560 & 83.10 & 690 & 75.30 \\
440 & 49.20 & 570 & 76.30 & 700 & 83.40 \\
450 & 63.80 & 580 & 84.60 & 710 & 68.00 \\
460 & 66.50 & 590 & 80.10 & 720 & 84.50 \\
470 & 79.10 & 600 & 79.50 & 730 & 72.50 \\
480 & 74.60 & 610 & 80.60 & 740 & 31.00 \\
490 & 76.30 & 620 & 79.80 & 750 &  2.81 \\
500 & 73.40 & 630 & 80.30 & 760 &  0.71 \\
510 & 79.20 & 640 & 76.00 & 770 &  0.37 \\
520 & 83.00 & 650 & 81.20 & 780 &  0.13 \\
\enddata
\end{deluxetable}

\begin{deluxetable}{llll}
\tablecolumns{4}
\tablewidth{0pc}
\tablecaption{Albedos of Giant Planets.}\label{ta:albedos}
\tablehead{
\colhead{Planet} & \colhead{Geometric Albedo} & \colhead{Geometric Albedo} & 
\colhead{Bond Albedo} \\
\colhead{} & \colhead{MOST Bandpass} & \colhead{All Wavelengths} & \colhead{}
 }
\startdata
HD 209458b & $\leq$ 0.25 & --  & -- \\
\hline
Jupiter & 0.50 & 0.274 $\pm$ 0.013 & 0.343 $\pm$ 0.032 \\
Saturn & 0.47 & 0.242 $\pm$ 0.012 & 0.342 $\pm$ 0.030 \\
Uranus & 0.43 & 0.208 $\pm$ 0.048 & 0.290 $\pm$ 0.051 \\
Neptune & 0.38 & 0.25 $\pm$ 0.02 & 0.31 $\pm$ 0.04 \\
\enddata
\end{deluxetable}

\centerline{\bf{TABLE CAPTIONS}}

{\sc Table} 1. {The parameters of targets observed during
  observations of the HD 209458 field. Coordinates are given in
  (J2000) and all measurements are from the Simbad astronomical database.}

{\sc Table} 2. {Best fit parameters for the eclipse of HD 209458b.
  The confidence levels correspond to the 1,2 and 3 $\sigma$ levels
  accordingly. The first row gives the ranges for the fit of $y_2-y_1$
  in mmag.  The second row gives the corresponding flux ratio of the
  planet to the star for the MOST bandpass calculated using Equation
  \ref{eq:flux}.  The third row gives the geometric albedo $A_g$ using
  Equation \ref{eq:ag}.
}

{\sc Table} 3. {Transmission values for the MOST bandpass filter.
  A finer grid of values is available upon request.}

{\sc Table} 4. {Albedo of HD 209458b compared to albedos of the solar system
giant planets.  Solar system planet geometric albedo in the MOST
bandpass computed from \citet{kar98} and other solar system planet
albedos from Voyager, Pioneer, and ground-based measurements described
in \citet{con89}.} 

}

\clearpage

\begin{figure}
\begin{center}
\plotone{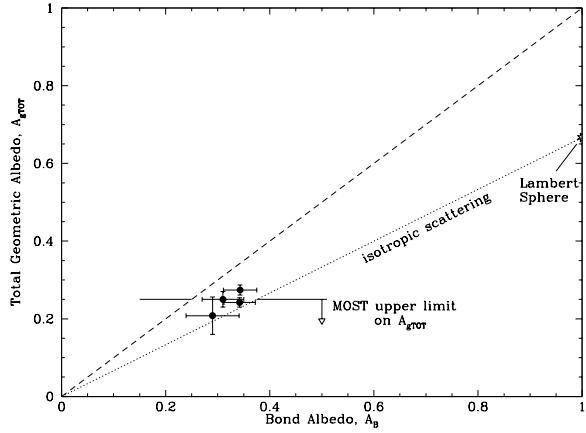}
\end{center}
\caption{
Relationship between the Bond albedo ($A_B$) and total
geometric albedo ($A_{g {\rm TOT}}$) for a Lambertian sphere and solar
system giant planets.  The points are for Uranus, Neptune, Saturn, and
Jupiter (in order of increasing $A_{B}$). The dotted line
($A_{g{\rm TOT}}/A_B$ = 0.67) is for Lambertian isotropic reflectance
(i.e., constant for all angles of incidence). The dashed line is the line
of equivalence where $A_{g{\rm TOT}}=A_B$ (all gas giant planets with
deep atmospheres must lie to its right). The wedge between the dotted and 
dashed lines defines a useful limiting region: it bounds the photometric
properties of most spherical bodies with deep atmospheres (with, e.g.,
Rayleigh scattering, clouds, dust, etc.). Hence for HD 209458b, with
the assumption of $A_{g{\rm TOT}}=A_g{\rm MOST}$ and $A_{g{\rm
    MOST}}\leq 0.25$, we estimate that $A_{\rm B}\leq 0.375$.
\label{fig:alb}}
\end{figure}

\begin{figure}
\begin{center}
\plotone{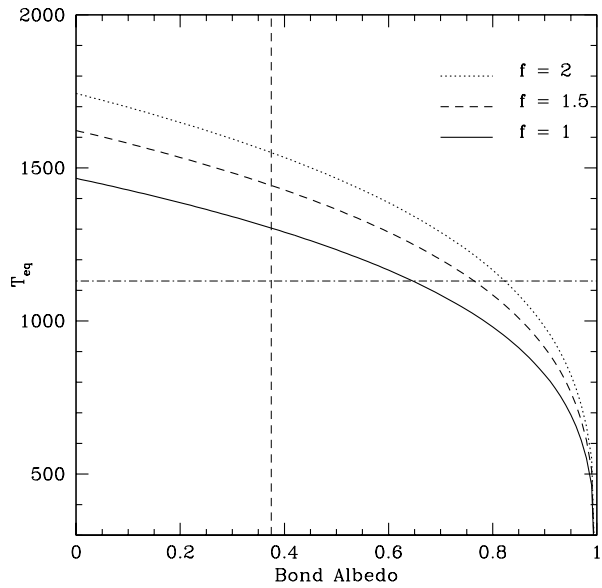}
\end{center}
\caption{The dayside $T_{eq}$ for HD209458b as a function of $A_B$ for
different values of $f$ (see equation~16). The approximate estimate of
$A_B$ is shown as a vertical dashed line.  The 24 $\mu$m brightness
temperature of 1130~K is shown (dash-dot line)\citep{dem05a} and can
be considered a lower limit to $T_{eq}$ \citep{sea05}. The upper left
quadrant is the  parameter space range for HD209458b based on our 1
$\sigma$ limits.
\label{fig:teq}}
\end{figure}

\end{document}